\documentclass{article}

\usepackage{arxiv}

\usepackage[utf8]{inputenc} 
\usepackage[T1]{fontenc}    
\usepackage{hyperref}       
\usepackage{url}            
\usepackage{booktabs}       
\usepackage{amsfonts}       
\usepackage{nicefrac}       
\usepackage{microtype}      
\usepackage{lipsum}
\usepackage{graphicx}
\usepackage{enumitem}
\usepackage{amsmath}
\usepackage{algorithm}
\usepackage[noend]{algpseudocode}
\usepackage{multirow}
\usepackage{natbib}
\makeatletter
\def\BState{\State\hskip-\ALG@thistlm}
\makeatother

\title{ROBUST BGA VOID DETECTION USING MULTI DIRECTIONAL SCAN ALGORITHMS}

\author{
  Vikas Ahuja\\
  Intel Corporation (India) Pvt. Ltd., \\
  Bangalore, India\\
  \texttt{vikas.ahuja@intel.com} \\
   \And
 Vijay Kumar Neeluru\\
  \texttt{vijaykumar413@gmail.com} \\
}

\begin{document}
\maketitle

\footnotetext[1]{This work was done while the second author was working for Intel India, Bangalore as a contractor.}

\begin{abstract}
The life time of electronic circuit’s board are impacted by the voids present in soldering balls. The quality inspection of 
solder balls by detecting and measuring the void is important to improve the board yield issues in electronic circuits. 
In general, the inspection is carried out manually, based on 2D or 3D X-ray images. For high quality inspection, it is 
difficult to detect and measure voids accurately with high repeatability through the manual inspection and it is time 
consuming process. In need of high quality and fast inspection, various approaches were proposed for void detection. But, 
lacks in robustness in dealing with various challenges like vias, reflections from the plating or vias, inconsistent lighting, 
noise, void-like artefacts, various void shapes, low resolution images and scalability to various devices. Robust BGA void 
detection becomes quite difficult problem, especially if the image size is very small (say, around 40x40) and with low contrast 
between void and the BGA background (say around 7 intensity levels on a scale of 255).  In this work, we propose novel approach 
for void detection based on the multi directional scanning. The proposed approach is able to segment the voids for low resolution 
images and can be easily scaled to various electronic manufacturing products. 
\end{abstract}


\section{Introduction}
In the electronic circuits manufacturing, a type of surface-mount named soldering ball grid array (BGA) is 
being used widely. Due to various reasons the voids will appear in these soldering balls that will reduce 
the life time of the device. Inspecting the quality of BGA to confirm the presence of voids is vital for quality 
inspection and to reduce the cost of manufacturing and increase the life time of device.~\cite{Hill2011}. 
In general, 2D and 3D X-ray imaging is used and voids are segmented based on the intensity differences within the BGA. 
Due to the low contrast between the voids and the BGA, it is difficult to accurately segment the void and an human expert is 
needed for the reliable inspection. The issue with the human inspection is the low repeatability due to the perceptual differences. 
As the number of soldering balls in BGA ranges from 10s to 100s, each ball is to be inspected in serial fashion resulting in 
longer inspection time. Due to these challenges in manual inspection, there is a need for robust void detection approach that 
can be scaled to all 2D X-ray acquisition device type, product type, layout of BGA and the void characteristics. 

The above technique is helping 
in more robust void detection even for low contrast small size BGA images.
  
Image processing techniques are usually applied for void detection by segmenting each soldering ball and 
segmenting the void regions within each soldering ball. The arrangement, size, intensity and number of soldering balls 
will vary for different 
devices. The challenging factors in void segmentation are relative intensity of voids, vias, plated-through holes, 
reflections from the plating or vias, inconsistent lighting, background traces, noise, void-like artifacts, and 
parallax effects. There is a need of robust techniques that are able to segment soldering balls and voids considering
all the mentioned factors.
Although many techniques are proposed since many years, but lacks in the robustness in dealing these various challenges.
The drawbacks of these approaches are scalability to new device types or different void 
characteristics and the parameters has to be tuned manually for different device and material types. 

In ~\cite{Said2012}, Laplacian of Gaussian (LoG) is applied to detect edges, via extraction, filtering out false voids 
and finally detect real voids. In ~\cite{Peng2012}, blob filters with various sizes are applied to segment the voids
of various sizes. In 
~\cite{Mouri2014}, void detection is formulated as the matrix decomposition problem by assuming that void as sparse 
component and non-negative matrix factorization approach is used to separate the void region from soldering ball. 

The end to end void inspection process can be broadly split into two steps. In the first step, the soldering balls are 
segmented from the input image and in the second step, the voids are to be segmented and analysed for each soldering 
ball. 

\textbf{Soldering Ball Segmentation}

In the first step, thresholding or background subtraction techniques are usually applied to remove the background, 
that works well to separate out the soldering balls from background. Some of the soldering balls are 
not detected due to shadowing of other components. In these cases, reference ball based matching technique is 
applied in ~\cite{Said2010}. Where, the reference template is used to match all the locations in the neighbourhood 
of occluded soldering ball. The best matched location is considered as the location of occluded one. In general, 
the pattern or arrangement of soldering balls will vary for each manufacturing device, prior information about the 
soldering balls pattern can simplify the soldering ball segmentation. To solve the pattern issue various approaches 
are proposed in ~\cite{Said2010}.   
 
\textbf{Void Detection} 
   
The challenging factors in void segmentation are relative intensity of voids, shape of void, vias, plated-through holes, 
reflections from the plating or vias, inconsistent lighting, background traces, noise, void-like artifacts, and 
parallax effects. The non-void soldering balls are of uniform intensity and voids are brighter with respect to 
intensity of non-void soldering ball. 

Various approaches are proposed in the literature to segment the voids. In ~\cite{Said2010}, multiple tools are applied to deal with 
the vias and false voids. In ~\cite{Peng2012}, blob filters with various sizes are applied to segment the voids of various sizes. 
In ~\cite{Mouri2014}, void detection is formulated as the matrix decomposition problem by assuming that void as sparse component 
and non-negative matrix factorization approach is used to separate the void region out of the BGA.    

The shape of voids are assumed as circular in ~\cite{Said2010} to simplify the detection. 
But, in practical, the voids can be of regular and irregular shapes and the overlapping of multiple regular voids will look as 
irregular void. The issue with the irregular shapes is 
difficulty in distinguishing the irregular shape voids and overlapping voids. Normally, multiple voids will 
appear. In some cases, the voids are partially or totally obscured by vias and the characteristics 
of via reflections are similar to those of the actual voids. The challenges are the scalable robust void segmentation 
irrespective of shape, number of voids, overlapping between multiple voids, and noises.  

In some manufacturing devices, the size of the soldering ball is less (at most 40x40) and the captured images are noisy.
Although various approaches are proposed, but lacks the robustness in detecting the voids. Common approach is to join 
the opened contours in detected Laplacian of Gaussian (LoG) edges by considering the nearest open edges. In the case of 
noisy images, low resolution soldering ball images the noisy edges will be appearing regularly. In this scenario, it is 
difficult to find the real voids by joining the nearest edges. The noisy images will create many open edges and difficult 
to find the real edges to join. The edge detection based approach doesn't consider the region characteristics to find the 
real voids, as the edges only consider the dissimilarity information between the neighbors. 

To improve the void detection, we propose the novel approach by considering the regional information by applying multi 
directional scanning algorithms. The LoG detected edges are scanned in radial direction for processing to find the 1D void 
regions and the 1D void regions are connected to form 2D voids based on neighbors and intensity of voids. Finally, connected 
component is applied to get the individual voids. 

The paper is organized as follow. The proposed approach is 
described in section 2, experimental results are presented in section 3 and finally concluded in section 4.

\section{Proposed Approach}
\label{sec:headings}

\subsection{Segmentation of Soldering Balls}
\begin{figure}[H]
\centering
\includegraphics[width=0.7\textwidth, height = 0.5\textheight]{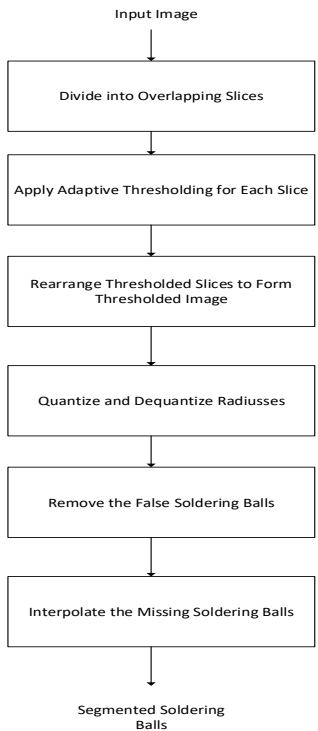}
\caption{Individual BGA Segmentation from Multi BGA Image}
\label{fig:BGA_segmentation}
\end{figure}

Given an input image, ~\cite{Said2010} applied adaptive thresholding to detect soldering balls and interpolating the
 soldering balls shadowed by the other components. The template matching approach is used to detect the 
 soldering balls shadowed by other components. The reference soldering ball is used to match all the locations 
 in the occluded region. The location having the higher correlation is considered as the center of the occluded 
 soldering ball. In this work, we follow the similar steps with some modifications. Instead of considering entire 
 image for adaptive thresholding, adaptive thresholding is applied for each slice independently to overcome the 
 sensitivity of adaptive thresholding on non-uniform illumination and other components. We observed that after 
 thresholding, there are some extra detections around the circle. Circles detection technique is applied to improve 
 the robustness on these noises. Location based technique is applied to verify the detection and 
to interpolate the location of missing ones. If any one of them is not detected then the 
distance between the neighboring balls will be more compared to the reference distance. We interpolate the location 
of this using the information of neighboring soldering balls. 
The pseudo code of the proposed algorithm is described below and the flow chart is given in Fig. ~\ref{fig:BGA_segmentation}. 

\begin{algorithm}
\caption{Algorithm for Individual BGA Segmentation from the Multi BGA Image}\label{Sol_seg}
\begin{algorithmic}[1]
\State Given an input image (consisting of multiple BGAs), divide it into non overlapping slices of size 300x400
\State For each slice, apply adaptive thresholding to segment the soldering balls
\State Rearrange the segmented slices in the corresponding position in the image to get the segmented image
\State Given the segmented image, to segment the soldering balls circles are detected in the image and store the radiuses 
($C_{radius}$) and locations ($C_{locations}$) of detected circular regions 
\State Quantize the radiuses with bin width of 4 and apply inverse quantization to get the ($C_{quantized_radius}$)
\State Find the mode of the quantized radiuses ($C_{quantized_mode}$), remove the circles which are not equal to the mode ($C_{quantized_mode}$)
\State Group the soldering balls based on the vertical alignment of locations. The soldering balls in each group are sorted 
according to the horizontal position and find the difference between the neighboring ones. The difference gives the distance 
between the neighboring soldering balls aligned in vertical position. If any soldering ball is missing in detection then the 
difference is more compared to the reference distance. In this case, the mis-detected soldering ball location is to be interpolated
\State The reference distance ($d_{ref}$) is the minimum distance between the neighbor ones. If the distance between the neighboring ones 
is greater (3*$d_{ref}$/2) then there is a missing soldering ball and it is interpolated using neighbors. 
The average of neighbor’s location is the location and average of radiuses is the radius 
\State In complex cases where the soldering balls are aligned in particular direction (apart from horizontal or vertical), 
a general approach is needed considering the alignment angle

\end{algorithmic}
\end{algorithm}

\subsection{Void Detection}

\begin{figure}[H]
\centering
\includegraphics[width=1.0\textwidth, height = 0.5\textheight]{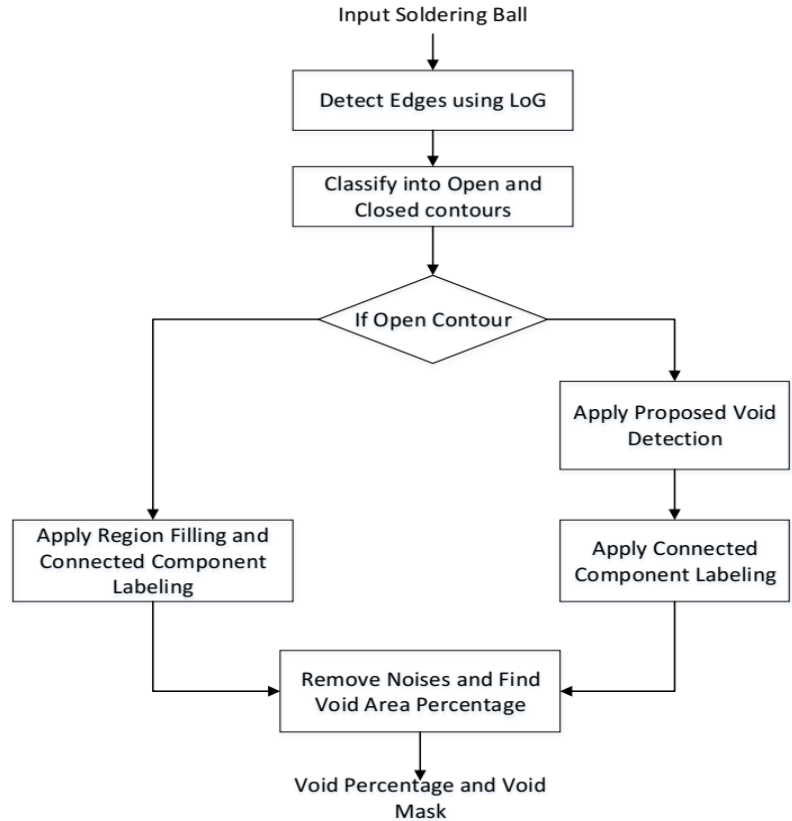}
\caption{Overall Algorithm for Void Detection}
\label{fig:Overall_algorithm}
\end{figure}

The overall flow for void detection is shown in Fig. ~\ref{fig:Overall_algorithm}

\subsubsection{Detect Edges}

After extracting soldering balls, the next step is to get the edges for each soldering ball. 
In this work, we apply LoG to detect the edges. Fig. ~\ref{fig:LoG_edge_mask} shows the detected edges using LoG for sample images. 

\subsubsection{Contours Classification}

\begin{figure}[H]
\centering
\includegraphics[width=0.5\textwidth, height = 0.3\textheight]{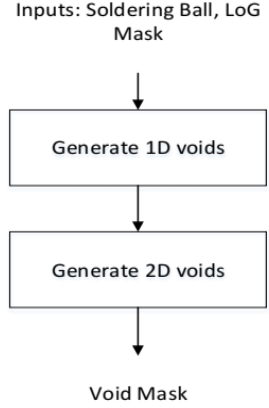}
\caption{Proposed Two Step Approach}
\label{fig:Proposed_two_step_approach}
\end{figure}

\begin{figure}[h]
\centering
\includegraphics[width=0.5\textwidth, height = 0.4\textheight]{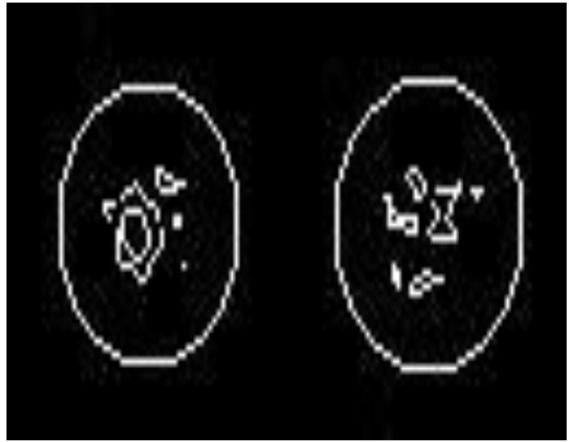}
\caption{LoG Edge Masks of Sample Images}
\label{fig:LoG_edge_mask}
\end{figure}

The edges corresponds to the dissimilarity between the two regions. In the case of noisy or low 
resolution soldering balls, it is possible to get many false positive edges. Fig. ~\ref{fig:LoG_edge_mask} shows the 
detected edges using LoG, and there is a high possibility of getting many false edges, open contours 
and it is difficult to filter out the false edges and close the contour to form the void region. In 
reference approaches ~\cite{Said2010}, the opened contours are joined to form closed contours by finding the nearest 
edge points. There are no explicit steps to filter the false edges and find the original void locations. 
In this work, the invalid edges are filtered out by considering the region information between this edge 
pixel and the next consecutive edge pixel scanned in particular order. The advantage of the proposed approach 
is filtering false edges and finding the 1D void regions can be done in a single step. 
Given the input image and LoG mask, the edges are divided into open and closer contours. For closed contours, 
regions are filled to get the void segment. The open contours are processed using the proposed method to form 
void regions. The void detection approach consists of two stages 1D Void generation and 2D void generation. The 
flow chart of the void approach is shown in the Fig. ~\ref{fig:Proposed_two_step_approach}.  

\subsubsection{Generate 1D Voids}

\begin{figure}[H]
\centering
\includegraphics[width=0.7\textwidth, height = 0.6\textheight]{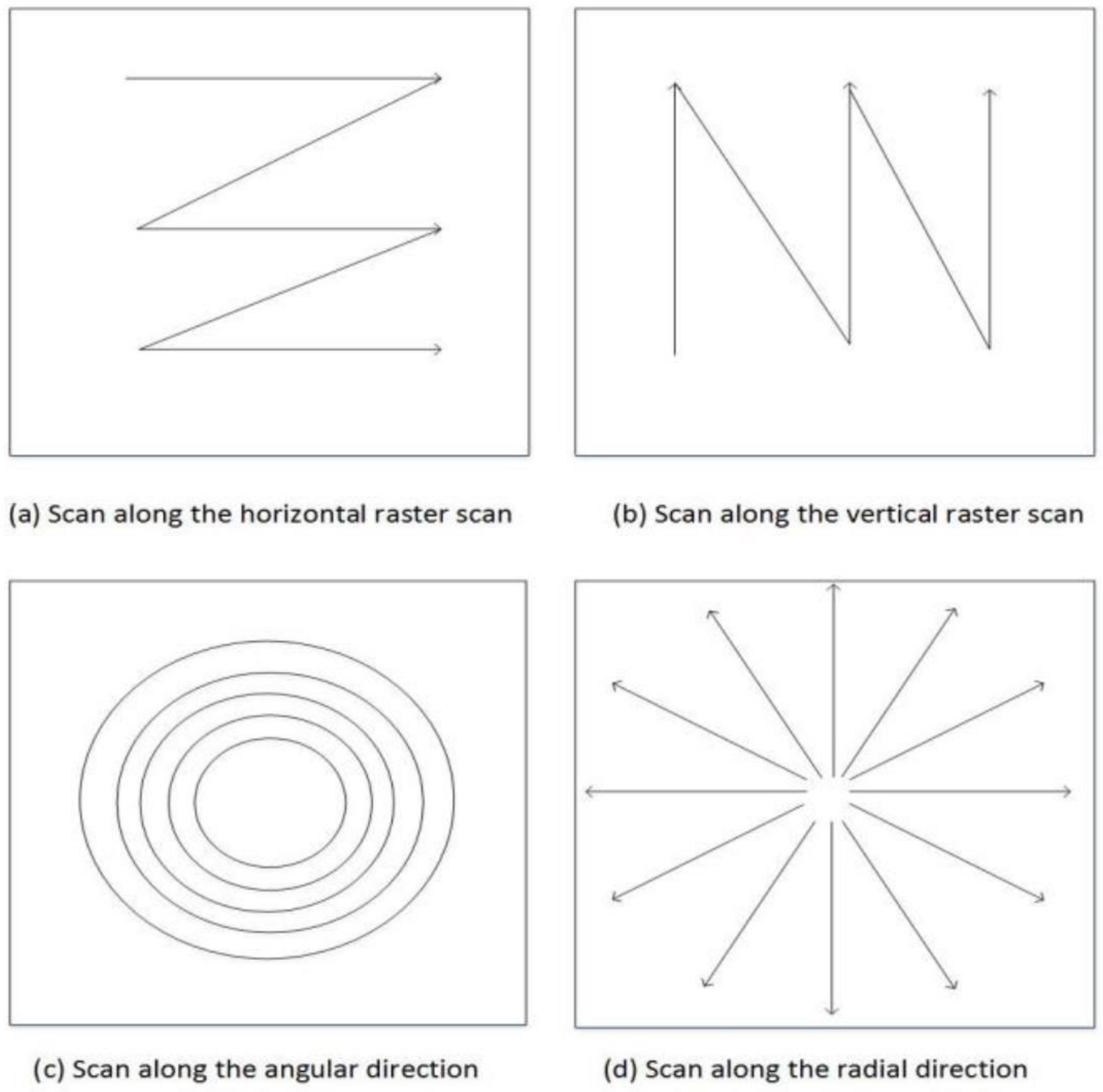}
\caption{Various Approaches to Scan the Pixel Locations for Void Detection}
\label{fig:Scanning}
\end{figure}

\begin{figure}[H]
\centering
\includegraphics[width=0.8\textwidth, height = 0.6\textheight]{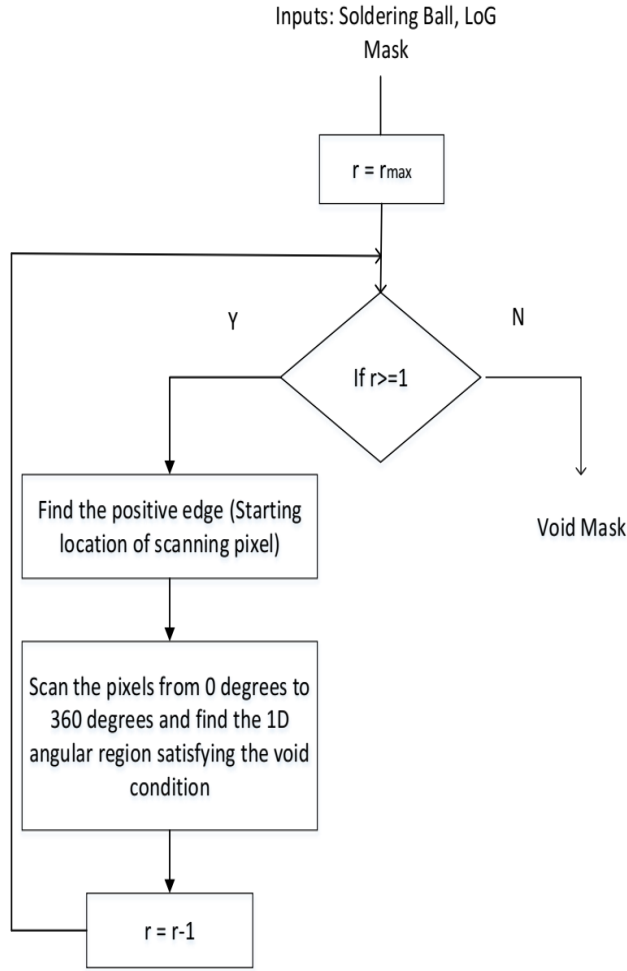}
\caption{Flow Chart for 1D Void Detection}
\label{fig:flowchart_1d_void}
\end{figure}

For each pixel belonging to the positive edge (background to void), the pixels are scanned in predefined order 
to find the negative edge (void to background). The pixels in scanned region are used to classify it whether 
the region belongs to void or background. The regular scanning techniques that can be used are horizontal, vertical, 
radial and angular directions as shown in Fig. ~\ref{fig:Scanning}. The core idea here is to explore the efficient scanning techniques 
for finding void.  Horizontal and vertical are the regular scanning techniques used in images. The alternative 
scanning approaches applied here are in radial $(r)$ and angular $(\beta)$ direction. 
The first advantage of scanning in $r$ and $\beta$ direction is the relatedness of circular scanning and the circular 
appearance of voids. In the case of low resolution soldering balls, it is difficult to distinguish the voids near 
the boundary. The intensity variations near the border are due to presence of void or due to edges of soldering balls. 
It is difficult to distinguish the void in this case using horizontal and vertical scanning. In circular and radial 
scanning, the intensity values are uniform at each radius r and the only intensity variations are due to void. In 
this case, $r$, $\beta$ is considered to be effective scanning approach compared to the horizontal and vertical scanning 
for void detection. For each radius from (r=$r_{max}$ to 1), pixels are scanned in angular direction for processing. 
rmax is the maximum radius of the soldering ball. The flow chart of the 1D void detection is shown in Fig. ~\ref{fig:flowchart_1d_void}.

At a particular radius r, cyclic shift is applied to find the pixel location which is a positive edge (transition 
from background to void). To do that, neighboring pixels ((r, $\beta$-1), (r, $\beta$)) are considered for checking the condition. 
If the intensity values at ((r, $\beta$-1), (r, $\beta$), (r, $\beta$+1)) are found to be in increasing order and (r, $\beta$) is an edge then 
edge at (r, $\beta$) is positive edge. The pixel at (r, $\beta$) is considered as starting location and the locations are scanned in 
angular direction until we find the negative edge (transition from void to background). The scanned locations which are 
in between the positive edge and negative edge are marked as void. Instead of blindly marking them as void pixels, one 
more extra condition that we applied is based on the maximum, minimum and average of intensity values of scanned locations. 
The intensity values in between the 
positive and negative edge are to be greater than background for a void. So, if the average value and the absolute 
value of maximum and minimum intensity difference are greater than ($thr_{1D}$=6) then the locations are marked as void. 
This procedure is repeated for each radius from r = $r_{max}$ to 1.   

\subsubsection{Generate 2D Voids}

\begin{figure}[H]
\centering
\includegraphics[width=1.0\textwidth, height = 0.7\textheight]{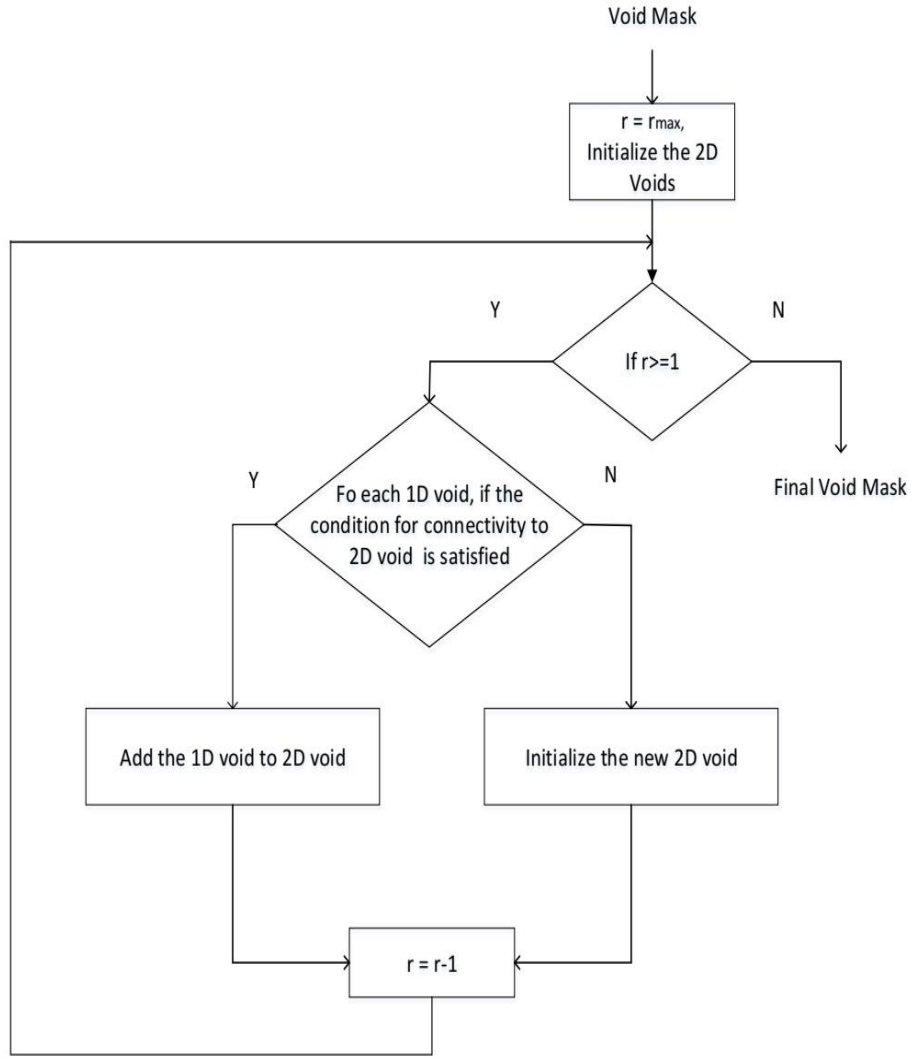}
\caption{Flow Chart for 2D Void Detection}
\label{fig:flowchart_2d_void}
\end{figure}

The detected partial voids (1D voids) belonging to same 2D void are to be assembled together. Fig. ~\ref{fig:flowchart_2d_void} shows the 
flow chart of 2D void detection. Starting from the maximum radius r = $r_{max}$, the 2D voids are initialized. The radius is decreased (r 
= $r_{max}$-1). Each 1D void in $r_{max}-1$ added to the one of the initialized 2D voids, if the connectivity condition is satisfied. 
New 2D void is created if the condition is not satisfied. The 2D voids are terminated, if none of the 1D void doesn’t satisfy 
the condition. The condition to check the connectivity is pixels belonging to 1D and 2D voids are to be 8-connected and the 
average intensity of 2D void and the 1D void has to be same. This process is repeated for all radiuses from $r_{max}$ to 1.  

\subsubsection{Filtering and Void Area Calculation}

After connecting the 1D voids to form 2D voids, connected component labelling is applied to separate the each void region. 
Regions whose area is less than minimum area ($A_{min}$ = 9 pixels) are filtered out to eliminate the noises. In 8-connectivity, 
the number of pixels covered with distances 1 and 2 are 1 and 9 pixels. Number of pixels 9 is considered as the minimum area 
for a valid void. The percentage of void is calculated based on the void percentage and the area of soldering ball. Void area 
is calculated by summing up number of pixels predicted as void. The area of soldering ball is calculated as the number of pixels 
inside the soldering ball. The percentage of void area with respect to soldering ball is calculated as 

Percentage of Void (\%) = (Area of void$/$Area of soldering ball) *100

\section{Experimental Results}

\begin{figure}[H]
\centering
\includegraphics[width=0.8\textwidth, height = 0.5\textheight]{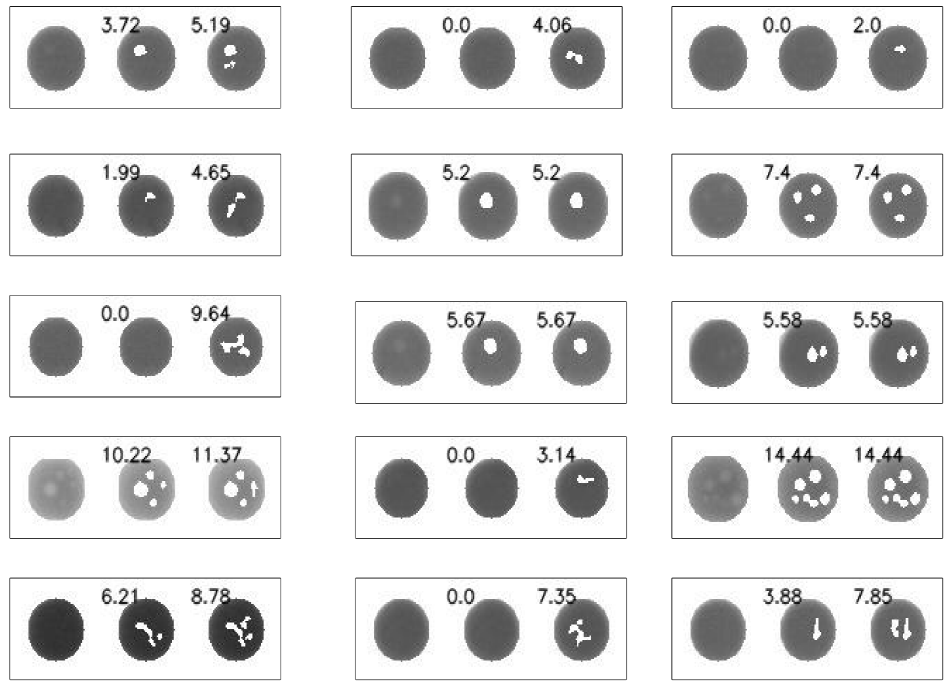}
\caption{Sample void detection results. The first image in each box represents input soldering ball. The second and third images represents the voids detected using reference and proposed}
\label{fig:void_detection_result}
\end{figure}

Fig. ~\ref{fig:void_detection_result} Shows the sample images of soldering balls and the detected voids using reference and proposed. 
The first image in each box represents the input image. The second and third represents the voids detected 
using reference and proposed. The numbers on the top of each image represents the void percentage. 
Compared to the reference approach, the proposed is able to detect more voids and no void will be missed. 
Due to the noisy edges, the reference approach is unable to fill the open contours. Whereas the proposed 
approach is able to scan using multidirectional scanning approach to form void regions. 

\section{Conclusion}

The quality inspection of BGA by measuring the voids is vital for board yield issues. Various traditional approaches 
were proposed for void segmentation but lacks in scalability and provides less accuracy. In this paper, we applied 
a novel approach for void segmentation by considering the regional information by applying multi 
directional scanning algorithms. The LoG detected edges are scanned in radial direction for processing to find the 1D void 
regions and the 1D void regions are connected to form 2D voids based on neighbors and intensity of voids. The proposed approach
is able to detect more voids compared to the reference approach.  

\bibliographystyle{unsrtnat}
\bibliography{references}

\end{document}